\documentclass[%
 aip,
 amsmath,amssymb,
 reprint,%
]{revtex4-1}

\usepackage{graphicx}
\usepackage{dcolumn}
\usepackage{bm}

\usepackage[utf8]{inputenc}
\usepackage[T1]{fontenc}
\usepackage{mathptmx}
\usepackage{etoolbox}

\makeatletter
\def\@email#1#2{%
 \endgroup
 \patchcmd{\titleblock@produce}
  {\frontmatter@RRAPformat}
  {\frontmatter@RRAPformat{\produce@RRAP{*#1\href{mailto:#2}{#2}}}\frontmatter@RRAPformat}
  {}{}
}%
\makeatother
\begin{document}

\preprint{AIP/123-QED}

\title[The ScIDEP Muon Radiography Project at the Egyptian Pyramid of Khafre]{}
\author{Manar Gamal}
 \affiliation{Institute of Basic and Applied Science, Egypt-Japan University of Science and Technology, Alexandria, Egypt.}
\author{Shereen Aly}
\affiliation{Institute of Basic and Applied Science, Egypt-Japan University of Science and Technology, Alexandria, Egypt.}
\author{Dora Geeraerts}
\affiliation{Inter-University Institute for High Energies, Vrije Universiteit Brussel, Brussels, Belgium.}

\author{Adam Hecht}
\affiliation{Department of Nuclear Engineering, University of New Mexico, Albuquerque, New Mexico, USA.}

\author{Richard Kouzes}
\affiliation{Pacific Northwest National Laboratory, Richland, Washington, USA.}
\email{rkouzes@ieee.org}

\author{Ayman Mahrouss}
\affiliation{Institute of Basic and Applied Science, Egypt-Japan University of Science and Technology, Alexandria, Egypt.}

\author{Isar Mostafanezhad}
\affiliation{Nalu Scientific, Honolulu, Hawaii, USA.}

\author{Alexandra Saftoiu}
\affiliation{Horia Hulubei National Institute for Physics and Nuclear Engineering, Magurele, Romania.}

\author{Denis Stanca}
\affiliation{Horia Hulubei National Institute for Physics and Nuclear Engineering, Magurele, Romania.}

\author{Michael Tytgat}
\email{michael.tytgat@vub.be}
\affiliation{Inter-University Institute for High Energies, Vrije Universiteit Brussel, Brussels, Belgium.}

\author{Catalin Vancea}
\affiliation{Horia Hulubei National Institute for Physics and Nuclear Engineering, Magurele, Romania.}

\author{Jesus Valencia}
\affiliation{Department of Nuclear Engineering, University of New Mexico, Albuquerque, New Mexico, USA.}

\author{Zhe Wang}
\affiliation{Inter-University Institute for High Energies, Vrije Universiteit Brussel, Brussels, Belgium.}

\date{\today}

\begin{abstract}
The ScIDEP Collaboration is constructing muon telescopes based on scintillator technology to investigate the internal structure of the Egyptian Pyramid of Khafre at Giza near Cairo using cosmic-ray muons. The collaboration aims to scan the pyramid from multiple viewpoints, both inside the king's burial chamber that is located centrally at the base of the pyramid, and outside of the pyramid, to potentially identify any new internal structures.  An overview of the project is presented, including the development of the data-acquisition system, the simulation framework, and very first detector studies. 
\end{abstract}

\maketitle

\section{\label{sec:intro}Introduction}

In muon radiography (or muography)~\cite{jimaging7120253}, muons produced in the collisions of cosmic rays with the earth's atmosphere are used as probes 
to study the internal structure of large objects such as buildings, industrial environments, containers and more. Cosmic-ray muons are generated with a broad range of energies, from MeV to TeV, and impinge on the Earth’s surface at sea level with about 160 muons/s/m$^{-2}$, with an angular distribution that follows closely a \begin{math}cos^2\end{math} law with respect to the vertical\cite{Hagmann}. At such energies, muons are minimally ionizing particles, and muons moving through materials undergo small angle scattering and attenuation proportional to the atomic number and density of the material. The attenuation feature is utilized in absorption muography, where the density integrated along the line of sight from a muon telescope to a target object can be determined by comparing the cosmic muon flux transmitted through the object with the flux from the open sky. A basic muography measurement produces a two-dimensional projection that may, by combining data from muon telescopes looking at the object from multiple locations, be used to create a three-dimensional reconstruction of a target volume.

Various research fields have used absorption muography measurements of density variation in ground-level or subsurface environments including, but not limited to, tunneling activities \cite{Kaiser, Guardincerri}, carbon sequestration \cite{Bonneville, Kudryavtsev}, non-proliferation and treaty verification \cite{Parker, Chatzidakis}, nuclear smuggling \cite{Bonomi, Checchia}, nuclear power reactor imaging \cite{Gomez}, oil and gas exploration and storage \cite{Kaiser}, mineral exploration \cite{Varga}, and various volcano studies~\cite{AGUbook}.

The application of muon radiography to archaeology spans several studies, including for example the imaging of the Pyramid of the Sun in Mexico \cite{Aguilar}, investigating the Palazzone necropolis Etruscan site in Perugia, Italy \cite{Borselli}, a study of early settlements of the city of Naples at Mt. Echia \cite{Cimmino}, imaging the Xi’an defensive walls in China \cite{Guorui}, and radiography of the Apollonia tumulus in Greece \cite{Marteau}. In fact, the first muography application to investigate Khafre’s pyramid at the Giza Plateau in Greater Cairo, Egypt appeared already in 1970 when the Physics Nobel Laureate L.W. Alvarez and his team looked for hidden chambers inside it using 4 m$^2$ spark chambers, but did not find any new structures \cite{Alvarez}. In 2017, a large void was discovered above the Grand Gallery in the Great Pyramid of Khufu \cite{Procureur}. Three different muon detectors were used for the pyramid scan: nuclear emulsions, scintillators, and micromegas. In 2023, the ScanPyramids project provided detailed data on the newly discovered North Face Corridor behind the North Face of the Khafre pyramid \cite{Morishima}.

The ScIDEP (Scintillator Imaging Detector for the Egyptian Pyramids) Collaboration~\cite{Aly} was formed to construct and operate muon telescopes to investigate the internal structure of the Egyptian Pyramid of Khafre at the Giza Plateau. 
Khafre’s pyramid has a slope of \begin{math}53^{o}10'\end{math}, an actual height of 136.4 m with a base length of 215.3 m, and is built out of limestone blocks weighing over 2 tons each \cite{Lehner}. Most of the original outer limestone casing was stripped over time. 
Although built just shortly after, during the Fourth Dynasty of the Old Kingdom of Ancient Egypt, and also just slightly smaller than the Great Pyramid of Khufu, the interior of Khafre's pyramid appears to be much simpler than that of the former, i.e. its known structure mainly consists of two entrance corridors on the north face, leading to a small subsidiary chamber and the larger central chamber, the King's burial chamber that measures about $14\times 5\times 6.8~$m$^3$. This might have been intentional by the builders, after having finished the construction of the Great Pyramid, or it might allow for hidden structures inside this pyramid that are yet to be discovered. Using the technique of absorption muography, the aim of ScIDEP is to scan this historical monument with muon telescopes placed at different viewpoints, such that in principle 3D information on its interior structure can be obtained. 

\section{\label{sec:scidep}ScIDEP Project Plan}

The ScIDEP Collaboration is developing two plastic scintillator-based muon tracker systems to be used to perform muography on the Pyramid of Khafre. The first is based on plain polyvinyl toluene (PVT) plates, while the second muon tracker is based on narrow polystyrene bars. In both cases, the scintillation light is collected with embedded wavelength shifting fibers that are connected to silicon photomultipliers (SiPM).

Once these detectors are operational and have been commissioned, the aim is to install at least one of them looking upward through the pyramid from inside the King’s burial chamber, which is suitably located at the bottom of the pyramid, slightly off-center from the central axis. The second detector could then be placed exterior to the pyramid, covering the top part of the monument plus part of the free sky. The precise configuration of the exterior detector in terms of positioning, size and angular coverage is still being investigated via simulation. 

\section{\label{sec:detector1}Detector 1: Polyvinyl Toluene Plates}

The initial version of the muon detector developed for these measurements consists of two EJ-200 (Eljen Technology) PVT scintillator plates measuring $61 \times 61 \times 2$~cm$^3$ each. Details on this detector design were previously described in \cite{Kouzes, Aly}. 

\begin{figure}
\includegraphics[width=0.48\textwidth]{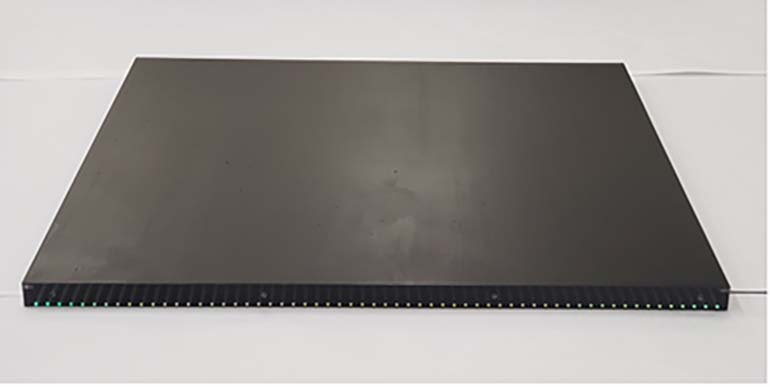}
\caption{\label{fig:detector1} One of the PVT plates showing the wavelength shifting fibers for one axis.}
\end{figure}

To enable the muon trajectory reconstruction, 2 mm diameter round BCF-92 (Saint-Gobain) wavelength shifting fibers are embedded orthogonally, spaced 1~cm apart, in both the surfaces of the PVT plates, yielding two times 60 fibers per PVT plane (see Figure \ref{fig:detector1}). The fibers are individually read out by S14160 (Hamamatsu) silicon photomultipliers (SiPMs), capturing the X-Y hit information. To suppress the internal reflection of the scintillation light,  which would impact the achievable position resolution, the surfaces of the PVT plates are painted black. While the effective position resolution of this initial configuration still has to be confirmed via lab measurements, our preliminary simulation studies indicate a position resolution better than 1~cm should be feasible.

The initial version of the data-acquisition system is based on CAEN PETIROC2A-ASIC boards supported by DT5550W modules. Custom-made readout electronics based on Weeroc ASICs in combination with FPGAs is under development to replace the CAEN system which has been difficult to operate. 

This detector is currently located at the Egypt-Japan University of Science and Technology (E-JUST) in Alexandria, Egypt, where the characterization of the system is ongoing.

\section{\label{sec:detector2}Detector 2: Polystyrene Bars}

\begin{figure}
\includegraphics[width=0.48\textwidth]{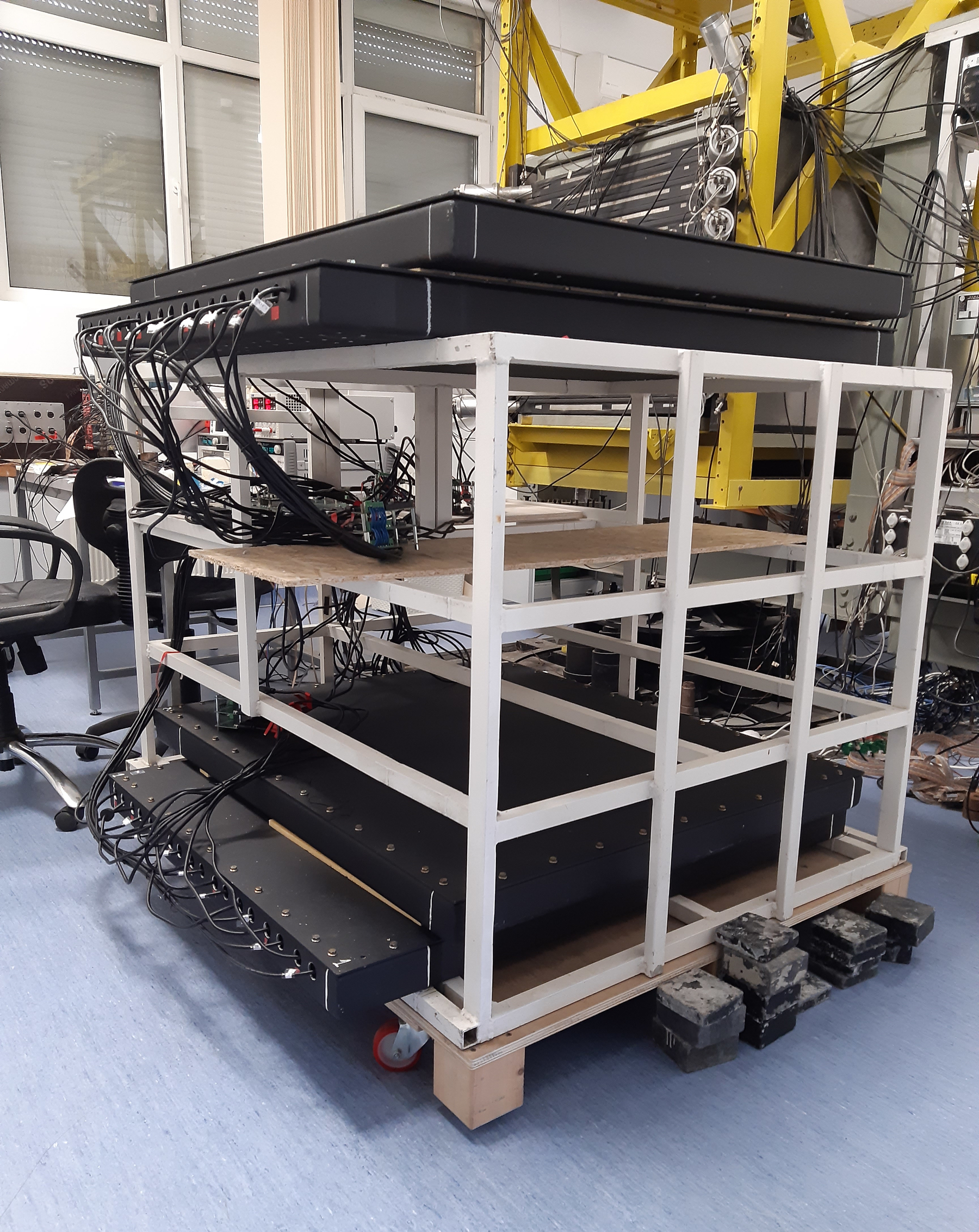}
\caption{\label{fig:mu36} The $\mu36$ detector in the laboratory.}
\end{figure}

\begin{figure}
\includegraphics[width=0.48\textwidth]{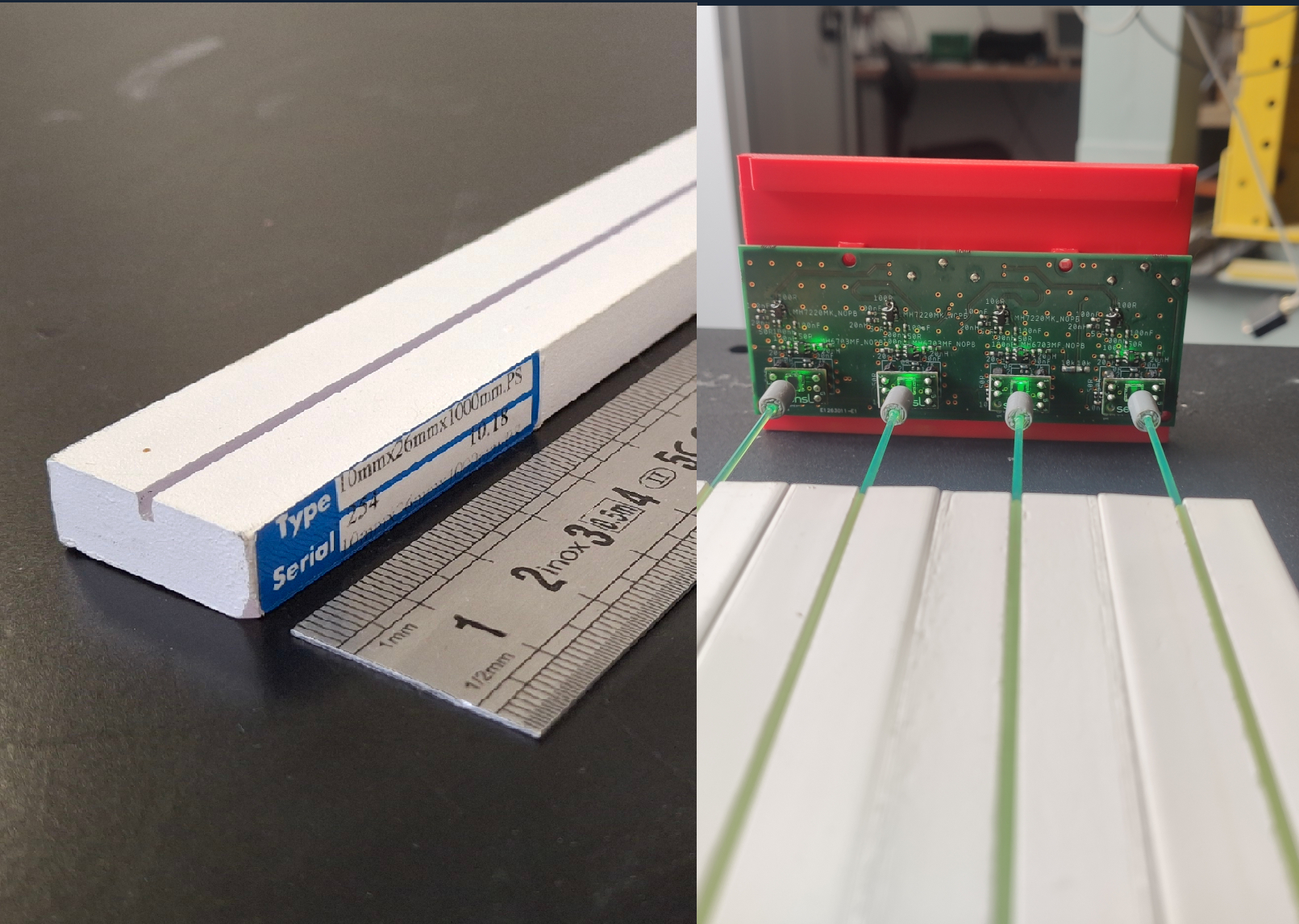}
\caption{\label{fig:mu36_comp} Scintillator bar (left) and optical fibres mounted inside the groove and connected to the SiPMs and the corresponding PCB (right).}
\end{figure}

The detector based on polystyrene bars is a muon telescope composed of four layers (see Figure~\ref{fig:mu36}). Each layer contains 36 scintillator bars read by wavelength shifting optical fibers and SiPM sensors. A scintillator bar, as well as the mounting of the fibers inside the grooves and their connection with the SiPMs, are shown in Figure \ref{fig:mu36_comp} without the covers necessary to ensure no light contamination from the outside or cross-talk between fibers. 
Consecutive layers of 36 bars each are placed perpendicular to one another to provide {\em x} and {\em y} coordinates of an incoming particle. The top two layers are distanced from the bottom two layers and a trajectory is reconstructed from the two points obtained.

The data acquisition is FPGA-based, using a Xilinx custom-made solution \cite{Gindac2021}. An event is considered valid and written to the output file when at least one bar has detected an event in each of the four detection layers, otherwise, trajectory reconstruction cannot be performed.

The scintillator bars show non-uniformities in response up to 50\% evidenced when testing separately. We assume this is an effect caused by casting a nonhomogeneous volume of plastic. To minimize the differences induced by the bars, we have grouped them according to their counting rates.
After polishing the fiber ends and mounting them in the scintillator grooves, they have been covered in PTFE (polytetrafluoroethylene) tape to ensure light containment.
Monte-Carlo simulations provide a 1.82~sr geometrical angular acceptance when the top layers are placed at a distance of 90 cm from the bottom layers, and 5.71~sr angular acceptance for all four plates stacked on top of each other.

\begin{figure}
\includegraphics[width=0.48\textwidth]{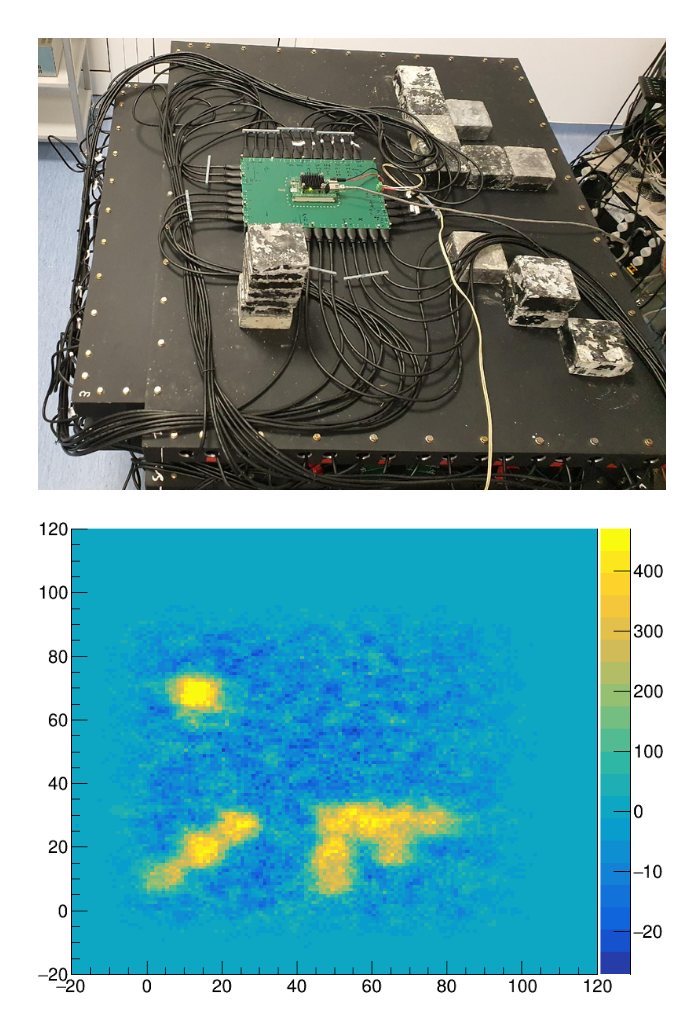}
\caption{Photo (top) and reconstructed image (bottom) of various shapes made of lead bricks for an exposure time of 2 days. 'Open-sky' data have been subtracted. \label{fig:36_labtest1} }
\end{figure}

\begin{figure}
\includegraphics[width=0.45\textwidth]{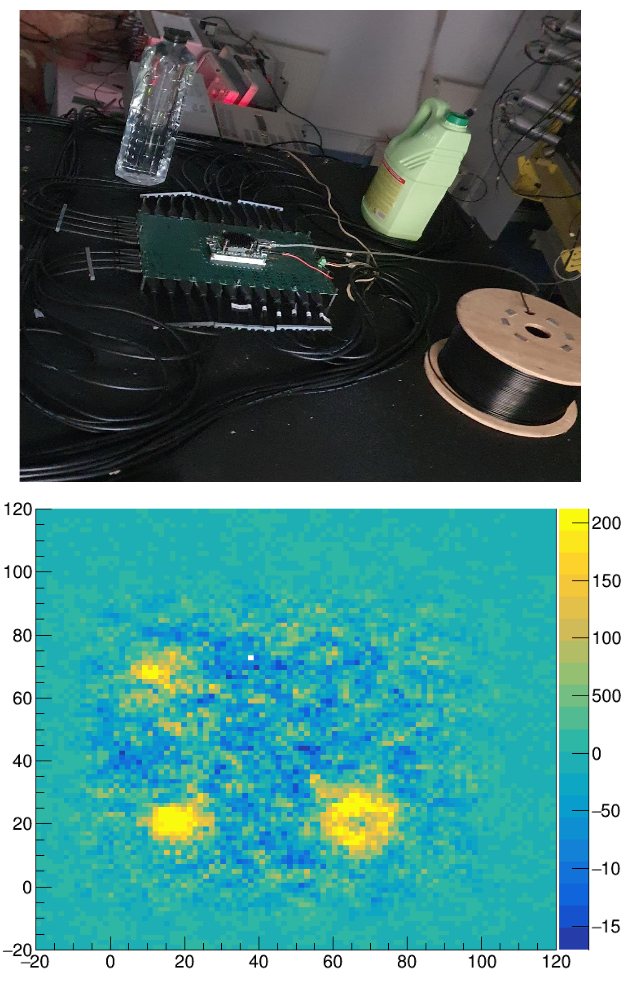}
\caption{Photo (top) and reconstructed image (bottom) containers of water and a cable roll for an exposure time of 2 days. 'Open-sky' data have been subtracted.\label{fig:36_labtest2} }
\end{figure}

Laboratory tests have been performed to assess the imaging capabilities of the detector. Figures \ref{fig:36_labtest1} and \ref{fig:36_labtest2} show photos of the measurements and the corresponding reconstructed images of various shapes and objects placed on top of the detector. Background data ('open-sky' muons) have been subtracted to show only the influence of the objects.

More details on the ${\mu36}$ detector can be found in \cite{Vancea}. The system is currently being prepared for {\em in situ} measurements in the Astroparticle Physics Laboratory of the Horia Hulubeei National Institute for Nuclear Physics and Engineering (IFIN-HH), Romania.

\section{\label{sec:level1}Simulations}

Simulations for the muon radiography measurements of the Khafre pyramid are ongoing with both Geant4 \cite{Agostinelli} and the Monte Carlo N-Particle (MCNP) \cite{MCNP} version 6. The previous simulation work under ScIDEP has been reported in \cite{Kouzes, Aly, Aly2}  Considering a simplified model of the pyramid, the results of the simulations showed that a modeled “hidden void” could be observed with \begin{math}10^8\end{math} emitted muons. The muon flux within the King’s chamber is expected to be about 1.44 muons/s/m$^{-2}$, with energies above 50 GeV that are required to be able to penetrate through the stone of the pyramid and reach the muon detection system.

A detailed Geant4-based simulation framework for the full project configuration is being implemented. It includes the detailed geometries of the detectors and a CAD model of the pyramid with its internal structure known today. The cosmic ray spectrum and distributions are derived from either the CRY \cite{Hagmann} or EcoMug \cite{Pagano} cosmic muon generators coupled to the Geant4 code. This simulation framework serves as input for the development of the muon data reconstruction chain and the muographic imaging algorithms. 

\section{\label{sec:conclusions}Summary}

The ScIDEP Collaboration is developing new muon radiography telescopes with the aim to study the internal structure of Khafre's Pyramid, the second largest pyramid at the Giza Plateau near Cairo, Egypt. The first muon telescope consists of two PVT scintillator plates with wavelength-shifting fibers in X-Y orientations, read out by SiPMs. This detector is currently being commissioned at E-JUST in Alexandria, Egypt. The second muon telescope has been developed in Magurele, Romania and is being prepared for delivery to Egypt. Once both systems have been commissioned locally in Egypt, preparations of the actual data taking at the pyramid will begin, including the logistics and administration to access the site.   
Meanwhile, a detailed simulation framework for the detectors and the setup has been developed using Geant4, and now the algorithms for the muon data reconstruction and muographic imaging are being implemented. 

\begin{acknowledgments}
Detector 1 was built through funds provided by the Egyptian Academy of Scientific Research and Technology (ASRT), Project ID: 6379. PNNL is operated for the US Department of Energy by Battelle under contract DE-AC05-76RLO 1830. Detector 2 was built with funds from the Romanian Ministry of Research, Innovation and Digitization, contract no. PN 23 21 01 02. The Science, Technology and Innovation Funding Authority (STDF) under grant number 48289 supported some of this work.
\end{acknowledgments}

\section*{Conflict of Interest}
The authors have no conflicts to disclose.

\section*{Data Availability Statement}

The data that support the findings of this study are available from the corresponding author upon reasonable request.

\nocite{*}
\bibliography{bibliog}

\providecommand{\noopsort}[1]{}\providecommand{\singleletter}[1]{#1}%
\begin{thebibliography}{31}%
\makeatletter
\providecommand \@ifxundefined [1]{%
 \@ifx{#1\undefined}
}%
\providecommand \@ifnum [1]{%
 \ifnum #1\expandafter \@firstoftwo
 \else \expandafter \@secondoftwo
 \fi
}%
\providecommand \@ifx [1]{%
 \ifx #1\expandafter \@firstoftwo
 \else \expandafter \@secondoftwo
 \fi
}%
\providecommand \natexlab [1]{#1}%
\providecommand \enquote  [1]{``#1''}%
\providecommand \bibnamefont  [1]{#1}%
\providecommand \bibfnamefont [1]{#1}%
\providecommand \citenamefont [1]{#1}%
\providecommand \href@noop [0]{\@secondoftwo}%
\providecommand \href [0]{\begingroup \@sanitize@url \@href}%
\providecommand \@href[1]{\@@startlink{#1}\@@href}%
\providecommand \@@href[1]{\endgroup#1\@@endlink}%
\providecommand \@sanitize@url [0]{\catcode `\\12\catcode `\$12\catcode `\&12\catcode `\#12\catcode `\^12\catcode `\_12\catcode `\%12\relax}%
\providecommand \@@startlink[1]{}%
\providecommand \@@endlink[0]{}%
\providecommand \url  [0]{\begingroup\@sanitize@url \@url }%
\providecommand \@url [1]{\endgroup\@href {#1}{\urlprefix }}%
\providecommand \urlprefix  [0]{URL }%
\providecommand \Eprint [0]{\href }%
\providecommand \doibase [0]{http://dx.doi.org/}%
\providecommand \selectlanguage [0]{\@gobble}%
\providecommand \bibinfo  [0]{\@secondoftwo}%
\providecommand \bibfield  [0]{\@secondoftwo}%
\providecommand \translation [1]{[#1]}%
\providecommand \BibitemOpen [0]{}%
\providecommand \bibitemStop [0]{}%
\providecommand \bibitemNoStop [0]{.\EOS\space}%
\providecommand \EOS [0]{\spacefactor3000\relax}%
\providecommand \BibitemShut  [1]{\csname bibitem#1\endcsname}%
\let\auto@bib@innerbib\@empty
\bibitem [{\citenamefont {Cimmino}(2021)}]{jimaging7120253}%
  \BibitemOpen
  \bibfield  {author} {\bibinfo {author} {\bibfnamefont {L.}~\bibnamefont {Cimmino}},\ }\bibfield  {title} {\enquote {\bibinfo {title} {Principles and perspectives of radiographic imaging with muons},}\ }\href {https://www.mdpi.com/2313-433X/7/12/253} {\bibfield  {journal} {\bibinfo  {journal} {Journal of Imaging}\ }\textbf {\bibinfo {volume} {7}} (\bibinfo {year} {2021})}\BibitemShut {NoStop}%
\bibitem [{\citenamefont {Hagmann}(2007)}]{Hagmann}%
  \BibitemOpen
  \bibfield  {author} {\bibinfo {author} {\bibfnamefont {C.}~\bibnamefont {Hagmann}},\ }\bibfield  {title} {\enquote {\bibinfo {title} {Cosmic-ray shower generator (cry) for monte carlo transport codes},}\ }\href@noop {} {\bibfield  {journal} {\bibinfo  {journal} {IEEE NSS Conf. Record}\ }\textbf {\bibinfo {volume} {2}},\ \bibinfo {pages} {1143–1146} (\bibinfo {year} {2007})}\BibitemShut {NoStop}%
\bibitem [{\citenamefont {Kaiser}(2018)}]{Kaiser}%
  \BibitemOpen
  \bibfield  {author} {\bibinfo {author} {\bibfnamefont {R.}~\bibnamefont {Kaiser}},\ }\bibfield  {title} {\enquote {\bibinfo {title} {Muography: overview and future directions},}\ }\href@noop {} {\bibfield  {journal} {\bibinfo  {journal} {Phil. Trans. R. Soc. A}\ }\textbf {\bibinfo {volume} {377}},\ \bibinfo {pages} {20180049} (\bibinfo {year} {2018})}\BibitemShut {NoStop}%
\bibitem [{\citenamefont {Guardincerri}\ \emph {et~al.}(2018)\citenamefont {Guardincerri} \emph {et~al.}}]{Guardincerri}%
  \BibitemOpen
  \bibfield  {author} {\bibinfo {author} {\bibfnamefont {E.}~\bibnamefont {Guardincerri}} \emph {et~al.},\ }\bibfield  {title} {\enquote {\bibinfo {title} {3d cosmic ray muon tomography from an underground tunnel},}\ }\href@noop {} {\bibfield  {journal} {\bibinfo  {journal} {Pure and App. Geophys.}\ }\textbf {\bibinfo {volume} {174}},\ \bibinfo {pages} {2133} (\bibinfo {year} {2018})}\BibitemShut {NoStop}%
\bibitem [{\citenamefont {Bonneville}\ \emph {et~al.}(2018)\citenamefont {Bonneville} \emph {et~al.}}]{Bonneville}%
  \BibitemOpen
  \bibfield  {author} {\bibinfo {author} {\bibfnamefont {A.}~\bibnamefont {Bonneville}} \emph {et~al.},\ }\bibfield  {title} {\enquote {\bibinfo {title} {Borehole muography of subsurface reservoirs},}\ }\href@noop {} {\bibfield  {journal} {\bibinfo  {journal} {Phil. Trans. R. Soc. A}\ }\textbf {\bibinfo {volume} {377}},\ \bibinfo {pages} {20180060} (\bibinfo {year} {2018})}\BibitemShut {NoStop}%
\bibitem [{\citenamefont {Kudryavtsev}\ \emph {et~al.}(2012)\citenamefont {Kudryavtsev} \emph {et~al.}}]{Kudryavtsev}%
  \BibitemOpen
  \bibfield  {author} {\bibinfo {author} {\bibfnamefont {V.~A.}\ \bibnamefont {Kudryavtsev}} \emph {et~al.},\ }\bibfield  {title} {\enquote {\bibinfo {title} {Monitoring subsurface co2 emplacement and security of storage using muon tomography},}\ }\href@noop {} {\bibfield  {journal} {\bibinfo  {journal} {Int. J. Greenhouse Gas Cont.}\ }\textbf {\bibinfo {volume} {11}},\ \bibinfo {pages} {21} (\bibinfo {year} {2012})}\BibitemShut {NoStop}%
\bibitem [{\citenamefont {Parker}\ \emph {et~al.}(2015)\citenamefont {Parker} \emph {et~al.}}]{Parker}%
  \BibitemOpen
  \bibfield  {author} {\bibinfo {author} {\bibfnamefont {H.~M.~O.}\ \bibnamefont {Parker}} \emph {et~al.},\ }\bibfield  {title} {\enquote {\bibinfo {title} {The use of ionising radiation to image nuclear fuel: A review},}\ }\href@noop {} {\bibfield  {journal} {\bibinfo  {journal} {Prog. in Nuc. Energy}\ }\textbf {\bibinfo {volume} {85}},\ \bibinfo {pages} {297--318} (\bibinfo {year} {2015})}\BibitemShut {NoStop}%
\bibitem [{\citenamefont {Chatzidakis}(2020)}]{Chatzidakis}%
  \BibitemOpen
  \bibfield  {author} {\bibinfo {author} {\bibfnamefont {S.}~\bibnamefont {Chatzidakis}},\ }\bibfield  {title} {\enquote {\bibinfo {title} {Progress report on model development for the transport of aerosol through microchannels},}\ }\href@noop {} {\bibfield  {journal} {\bibinfo  {journal} {ORNL}\ }\textbf {\bibinfo {volume} {SPR-2020}},\ \bibinfo {pages} {1728} (\bibinfo {year} {2020})}\BibitemShut {NoStop}%
\bibitem [{\citenamefont {Bonomi}\ \emph {et~al.}(2020)\citenamefont {Bonomi} \emph {et~al.}}]{Bonomi}%
  \BibitemOpen
  \bibfield  {author} {\bibinfo {author} {\bibfnamefont {G.}~\bibnamefont {Bonomi}} \emph {et~al.},\ }\bibfield  {title} {\enquote {\bibinfo {title} {Applications of cosmic-ray muons},}\ }\href@noop {} {\bibfield  {journal} {\bibinfo  {journal} {Prog. in Part. and Nuc. Phys.}\ }\textbf {\bibinfo {volume} {112}},\ \bibinfo {pages} {103768} (\bibinfo {year} {2020})}\BibitemShut {NoStop}%
\bibitem [{\citenamefont {Checchia}(2016)}]{Checchia}%
  \BibitemOpen
  \bibfield  {author} {\bibinfo {author} {\bibfnamefont {P.}~\bibnamefont {Checchia}},\ }\bibfield  {title} {\enquote {\bibinfo {title} {Review of possible applications of cosmic muon tomography},}\ }\href@noop {} {\bibfield  {journal} {\bibinfo  {journal} {JINST}\ }\textbf {\bibinfo {volume} {11}},\ \bibinfo {pages} {C12072} (\bibinfo {year} {2016})}\BibitemShut {NoStop}%
\bibitem [{\citenamefont {Gomez}(2019)}]{Gomez}%
  \BibitemOpen
  \bibfield  {author} {\bibinfo {author} {\bibfnamefont {H.}~\bibnamefont {Gomez}},\ }\bibfield  {title} {\enquote {\bibinfo {title} {Muon tomography using micromegas detectors: From archaeology to nuclear safety applications},}\ }\href@noop {} {\bibfield  {journal} {\bibinfo  {journal} {NIM A}\ }\textbf {\bibinfo {volume} {936}},\ \bibinfo {pages} {14} (\bibinfo {year} {2019})}\BibitemShut {NoStop}%
\bibitem [{\citenamefont {Varga}\ \emph {et~al.}(2020)\citenamefont {Varga} \emph {et~al.}}]{Varga}%
  \BibitemOpen
  \bibfield  {author} {\bibinfo {author} {\bibfnamefont {D.}~\bibnamefont {Varga}} \emph {et~al.},\ }\bibfield  {title} {\enquote {\bibinfo {title} {Detector developments for high performance muography applications},}\ }\href@noop {} {\bibfield  {journal} {\bibinfo  {journal} {NIM A}\ }\textbf {\bibinfo {volume} {958}},\ \bibinfo {pages} {162236} (\bibinfo {year} {2020})}\BibitemShut {NoStop}%
\bibitem [{\citenamefont {Oláh}, \citenamefont {Tanaka},\ and\ \citenamefont {Varga}(2022)}]{AGUbook}%
  \BibitemOpen
  \bibinfo {editor} {\bibfnamefont {L.}~\bibnamefont {Oláh}}, \bibinfo {editor} {\bibfnamefont {H.~K.~M.}\ \bibnamefont {Tanaka}}, \ and\ \bibinfo {editor} {\bibfnamefont {D.}~\bibnamefont {Varga}},\ eds.,\ \href {\doibase 10.1002/9781119722748} {\emph {\bibinfo {title} {Muography: Exploring Earth's Subsurface with Elementary Particles}}}\ (\bibinfo  {publisher} {American Geophysical Union},\ \bibinfo {year} {2022})\BibitemShut {NoStop}%
\bibitem [{\citenamefont {Aguilar}\ \emph {et~al.}(2013)\citenamefont {Aguilar} \emph {et~al.}}]{Aguilar}%
  \BibitemOpen
  \bibfield  {author} {\bibinfo {author} {\bibfnamefont {S.}~\bibnamefont {Aguilar}} \emph {et~al.},\ }\bibfield  {title} {\enquote {\bibinfo {title} {Search for cavities in the teotihuacan pyramid of the sun using cosmic muons: preliminary results},}\ }\href@noop {} {\bibfield  {journal} {\bibinfo  {journal} {Proc. 33rd ICRC}\ ,\ \bibinfo {pages} {0364}} (\bibinfo {year} {2013})}\BibitemShut {NoStop}%
\bibitem [{\citenamefont {Borselli}\ \emph {et~al.}(2024)\citenamefont {Borselli} \emph {et~al.}}]{Borselli}%
  \BibitemOpen
  \bibfield  {author} {\bibinfo {author} {\bibfnamefont {D.}~\bibnamefont {Borselli}} \emph {et~al.},\ }\bibfield  {title} {\enquote {\bibinfo {title} {Muographic study of the palazzone necropolis (perugia-italy)},}\ }\href@noop {} {\bibfield  {journal} {\bibinfo  {journal} {JAIS}\ }\textbf {\bibinfo {volume} {467}} (\bibinfo {year} {2024})}\BibitemShut {NoStop}%
\bibitem [{\citenamefont {Cimmino}\ \emph {et~al.}(2019)\citenamefont {Cimmino} \emph {et~al.}}]{Cimmino}%
  \BibitemOpen
  \bibfield  {author} {\bibinfo {author} {\bibfnamefont {L.}~\bibnamefont {Cimmino}} \emph {et~al.},\ }\bibfield  {title} {\enquote {\bibinfo {title} {3d muography for the search of hidden cavities},}\ }\href@noop {} {\bibfield  {journal} {\bibinfo  {journal} {Sci. Rep.}\ }\textbf {\bibinfo {volume} {9}},\ \bibinfo {pages} {2974} (\bibinfo {year} {2019})}\BibitemShut {NoStop}%
\bibitem [{\citenamefont {Guorui}\ \emph {et~al.}(2023)\citenamefont {Guorui} \emph {et~al.}}]{Guorui}%
  \BibitemOpen
  \bibfield  {author} {\bibinfo {author} {\bibfnamefont {L.}~\bibnamefont {Guorui}} \emph {et~al.},\ }\bibfield  {title} {\enquote {\bibinfo {title} {High-precision muography in archaeogeophysics: A case study on xi’an defensive walls},}\ }\href@noop {} {\bibfield  {journal} {\bibinfo  {journal} {J. Appl. Phys.}\ }\textbf {\bibinfo {volume} {133}},\ \bibinfo {pages} {014901} (\bibinfo {year} {2023})}\BibitemShut {NoStop}%
\bibitem [{\citenamefont {Marteau}\ \emph {et~al.}(2018)\citenamefont {Marteau} \emph {et~al.}}]{Marteau}%
  \BibitemOpen
  \bibfield  {author} {\bibinfo {author} {\bibfnamefont {J.}~\bibnamefont {Marteau}} \emph {et~al.},\ }\bibfield  {title} {\enquote {\bibinfo {title} {Applied muography : from volcanology to archaelogy with a mobile muon detector (diaphane / arch\'e)},}\ }\href@noop {} {\bibfield  {journal} {\bibinfo  {journal} {https://ui.adsabs.harvard.edu/abs/2018AGUFMNS23B0705M}\ } (\bibinfo {year} {2018})}\BibitemShut {NoStop}%
\bibitem [{\citenamefont {Alvarez}(1970)}]{Alvarez}%
  \BibitemOpen
  \bibfield  {author} {\bibinfo {author} {\bibfnamefont {L.}~\bibnamefont {Alvarez}},\ }\bibfield  {title} {\enquote {\bibinfo {title} {Search for hidden chambers in the pyramids},}\ }\href@noop {} {\bibfield  {journal} {\bibinfo  {journal} {Science}\ }\textbf {\bibinfo {volume} {167}},\ \bibinfo {pages} {832–839} (\bibinfo {year} {1970})}\BibitemShut {NoStop}%
\bibitem [{\citenamefont {Procureur}\ \emph {et~al.}(2023)\citenamefont {Procureur} \emph {et~al.}}]{Procureur}%
  \BibitemOpen
  \bibfield  {author} {\bibinfo {author} {\bibfnamefont {S.}~\bibnamefont {Procureur}} \emph {et~al.},\ }\bibfield  {title} {\enquote {\bibinfo {title} {Precise characterization of a corridor-shaped structure in khufu’s pyramid by observation of cosmic-ray muons},}\ }\href@noop {} {\bibfield  {journal} {\bibinfo  {journal} {Nature Communications}\ }\textbf {\bibinfo {volume} {14}},\ \bibinfo {pages} {1144} (\bibinfo {year} {2023})}\BibitemShut {NoStop}%
\bibitem [{\citenamefont {Morishima}\ \emph {et~al.}(2017)\citenamefont {Morishima} \emph {et~al.}}]{Morishima}%
  \BibitemOpen
  \bibfield  {author} {\bibinfo {author} {\bibfnamefont {K.}~\bibnamefont {Morishima}} \emph {et~al.},\ }\bibfield  {title} {\enquote {\bibinfo {title} {Discovery of a big void in khufu’s pyramid by observation of cosmic-ray muons},}\ }\href@noop {} {\bibfield  {journal} {\bibinfo  {journal} {Nature}\ }\textbf {\bibinfo {volume} {552}},\ \bibinfo {pages} {386--390} (\bibinfo {year} {2017})}\BibitemShut {NoStop}%
\bibitem [{\citenamefont {Aly}\ \emph {et~al.}(2024)\citenamefont {Aly} \emph {et~al.}}]{Aly}%
  \BibitemOpen
  \bibfield  {author} {\bibinfo {author} {\bibfnamefont {S.}~\bibnamefont {Aly}} \emph {et~al.},\ }\bibfield  {title} {\enquote {\bibinfo {title} {The scidep project at the egyptian pyramid of khafre},}\ }\href@noop {} {\bibfield  {journal} {\bibinfo  {journal} {JAIS}\ }\textbf {\bibinfo {volume} {470}} (\bibinfo {year} {2024})}\BibitemShut {NoStop}%
\bibitem [{\citenamefont {Lehner}(1997)}]{Lehner}%
  \BibitemOpen
  \bibfield  {author} {\bibinfo {author} {\bibfnamefont {M.}~\bibnamefont {Lehner}},\ }\href@noop {} {\emph {\bibinfo {title} {The Complete Pyramids: Solving the Ancient Mysteries, 1st Ed. edition}}}\ (\bibinfo  {publisher} {Thames \& Hudson},\ \bibinfo {year} {1997})\BibitemShut {NoStop}%
\bibitem [{\citenamefont {Kouzes}\ \emph {et~al.}(2022)\citenamefont {Kouzes} \emph {et~al.}}]{Kouzes}%
  \BibitemOpen
  \bibfield  {author} {\bibinfo {author} {\bibfnamefont {R.~T.}\ \bibnamefont {Kouzes}} \emph {et~al.},\ }\bibfield  {title} {\enquote {\bibinfo {title} {Novel muon tomography detector for the pyramids},}\ }\href@noop {} {\bibfield  {journal} {\bibinfo  {journal} {JAIS}\ }\textbf {\bibinfo {volume} {240}} (\bibinfo {year} {2022})}\BibitemShut {NoStop}%
\bibitem [{\citenamefont {Gindac}\ \emph {et~al.}(2021)\citenamefont {Gindac} \emph {et~al.}}]{Gindac2021}%
  \BibitemOpen
  \bibfield  {author} {\bibinfo {author} {\bibfnamefont {A.}~\bibnamefont {Gindac}} \emph {et~al.},\ }\bibfield  {title} {\enquote {\bibinfo {title} {Open-source soc-based off-the-shelf industrial-grade low-cost logic analyzer},}\ }\href@noop {} {\bibfield  {journal} {\bibinfo  {journal} {Romanian Journal of Information Science and Technology}\ }\textbf {\bibinfo {volume} {1}},\ \bibinfo {pages} {117--126} (\bibinfo {year} {2021})}\BibitemShut {NoStop}%
\bibitem [{\citenamefont {Vancea}\ \emph {et~al.}(2025)\citenamefont {Vancea} \emph {et~al.}}]{Vancea}%
  \BibitemOpen
  \bibfield  {author} {\bibinfo {author} {\bibfnamefont {C.}~\bibnamefont {Vancea}} \emph {et~al.},\ }\bibfield  {title} {\enquote {\bibinfo {title} {$\mu36$: a sipm-read scintillator detector designed for muography applications},}\ }\href@noop {} {\bibfield  {journal} {\bibinfo  {journal} {Accepted by JINST}\ } (\bibinfo {year} {2025})}\BibitemShut {NoStop}%
\bibitem [{\citenamefont {Agostinelli}\ \emph {et~al.}(2003)\citenamefont {Agostinelli} \emph {et~al.}}]{Agostinelli}%
  \BibitemOpen
  \bibfield  {author} {\bibinfo {author} {\bibfnamefont {S.}~\bibnamefont {Agostinelli}} \emph {et~al.},\ }\bibfield  {title} {\enquote {\bibinfo {title} {Geant4—a simulation toolkit},}\ }\href@noop {} {\bibfield  {journal} {\bibinfo  {journal} {NIM A}\ }\textbf {\bibinfo {volume} {506}},\ \bibinfo {pages} {3} (\bibinfo {year} {2003})}\BibitemShut {NoStop}%
\bibitem [{\citenamefont {Monte-Carlo-Team}(2017)}]{MCNP}%
  \BibitemOpen
  \bibfield  {author} {\bibinfo {author} {\bibnamefont {Monte-Carlo-Team}},\ }\href@noop {} {\enquote {\bibinfo {title} {Mcnp -- a general monte carlo n-particle transport code, version 6},}\ }\bibinfo {type} {Tech. Rep.}\ (\bibinfo  {institution} {Los Alamos National Laboratory},\ \bibinfo {address} {Los Alamos, NM, USA},\ \bibinfo {year} {2017})\BibitemShut {NoStop}%
\bibitem [{\citenamefont {Aly}\ \emph {et~al.}(2022)\citenamefont {Aly} \emph {et~al.}}]{Aly2}%
  \BibitemOpen
  \bibfield  {author} {\bibinfo {author} {\bibfnamefont {S.}~\bibnamefont {Aly}} \emph {et~al.},\ }\bibfield  {title} {\enquote {\bibinfo {title} {Simulation studies of a novel muography detector for the great pyramids},}\ }\href@noop {} {\bibfield  {journal} {\bibinfo  {journal} {JAIS}\ }\textbf {\bibinfo {volume} {306}} (\bibinfo {year} {2022})}\BibitemShut {NoStop}%
\bibitem [{\citenamefont {Pagano}\ \emph {et~al.}(2021)\citenamefont {Pagano} \emph {et~al.}}]{Pagano}%
  \BibitemOpen
  \bibfield  {author} {\bibinfo {author} {\bibfnamefont {D.}~\bibnamefont {Pagano}} \emph {et~al.},\ }\bibfield  {title} {\enquote {\bibinfo {title} {Ecomug: An efficient cosmic muon generator for cosmic-ray muon applications},}\ }\href@noop {} {\bibfield  {journal} {\bibinfo  {journal} {NIM A}\ }\textbf {\bibinfo {volume} {1014}},\ \bibinfo {pages} {165732} (\bibinfo {year} {2021})}\BibitemShut {NoStop}%
\bibitem [{\citenamefont {Tanaka}\ \emph {et~al.}(2007)\citenamefont {Tanaka} \emph {et~al.}}]{Tanaka}%
  \BibitemOpen
  \bibfield  {author} {\bibinfo {author} {\bibfnamefont {H.~K.~M.}\ \bibnamefont {Tanaka}} \emph {et~al.},\ }\bibfield  {title} {\enquote {\bibinfo {title} {High resolution imaging in the inhomogeneous crust with cosmic-ray muon radiography: The density structure below the volcanic crater floor of mt. asama, japan},}\ }\href@noop {} {\bibfield  {journal} {\bibinfo  {journal} {Earth and Planetary Sci. Let.}\ }\textbf {\bibinfo {volume} {263}},\ \bibinfo {pages} {104} (\bibinfo {year} {2007})}\BibitemShut {NoStop}%
\end{thebibliography}%

\end{document}